\documentclass{ifacconf}

\usepackage{graphicx}      
\usepackage{natbib}        

\usepackage{amsmath,amssymb,bm,mathrsfs,bbm}
\usepackage{bbm}
\usepackage{dsfont}

\usepackage{algorithm,algorithmic}
\usepackage{xcolor}

\usepackage{epstopdf}
\DeclareGraphicsExtensions{.pdf,.png,.jpg,.eps}              

\newenvironment{Proof}{\begin{pf}}{\end{pf}}
\newenvironment{lemma}{\begin{lem}}{\end{lem}}
\newenvironment{theorem}{\begin{thm}}{\end{thm}}
{\theorembodyfont{\rmfamily}}

\defcitealias{sec10b5}{SEC Rule 10b--5}
\defcitealias{euMAR17}{MAR Art.~17}

\begin{document}
\begin{frontmatter}

\title{\!\!\!Timely Information for Strategic Persuasion\thanksref{footnoteinfo}\!\!}

\thanks[footnoteinfo]{This work was supported by Tubitak 2232-B program (Project No:124C533).}

\author{Ahmet Bugra Gundogan\qquad}
\author{Melih Bastopcu}

\address{Department of Electrical and Electronics Engineering, Bilkent University, 06800 Ankara, T\"urkiye.\\
\texttt{bugra.gundogan@bilkent.edu.tr}, \texttt{bastopcu@bilkent.edu.tr}
}

\begin{abstract}

This work investigates a dynamic variant of a persuasion problem, in which a strategic sender seeks to influence a receiver's belief over time by controlling the timing of information disclosure, under resource constraints. We consider a binary information source (i.e., taking values 0 or 1), where the source's state evolves according to a continuous-time Markov chain (CTMC). In this setting, the receiver aims to estimate the source's state as accurately as possible. In contrast, the sender seeks to persuade the receiver to estimate the state to be 1, regardless of whether this estimate reflects the true state. This misalignment between their objectives naturally leads to a Stackelberg game formulation where the sender, acting as the leader, chooses an information-revelation policy, and the receiver, as the follower, decides whether to follow the sender’s messages. As a result, the sender's objective is to maximize the long-term average time that the receiver's estimate equals 1, subject to a total sampling constraint and a constraint for the receiver to follow the sender's messages called \emph{incentive compatibility (IC) constraint.}
We first consider the single-source problem and show that the sender’s optimal policy is to allocate the minimal sampling rate to the undesired state 0 (just enough to satisfy the IC constraint) and assign the remaining sampling rate to the desired state 1. Next, we extend the analysis to the multi-source case, where each source has a different minimal sampling rate. Our results show that the sender can leverage the timeliness of the revealed information to influence the receiver and thereby increase its own utility.    
\end{abstract}


\end{frontmatter}

\section{Introduction}\vspace{-0.07cm}
In public markets, \emph{misrepresentation is illegal}
\citepalias{sec10b5} while \emph{delaying truthful disclosure can be lawful} in specific cases \citepalias{euMAR17}. Inspired by this, we model a firm whose fundamentals switch between “favorable" (denoted as state 1) and “unfavorable" (denoted as state 0) regimes as a binary continuous-time Markov chain (CTMC). The sender (owners / management) cannot lie about outcomes, but can lawfully manage \emph{when} to release verifiable updates by communicating more frequently when fundamentals are good (with rate \(s\)) and more slowly when they are bad (with rate \(c\)) subject to an investor- / attention budget \(R\). As another example, we can consider a news provider that supports a campaign, but reports only true updates. The campaign’s “momentum" can be similarly modeled with binary unfavorable \((0)\) and favorable \((1)\) states. Here, the news provider cannot falsify content, but it can choose when to air it: it releases favorable news with a higher rate and unfavorable news with a slower rate, under a total airtime budget (so that timing of the news, not its content, is the instrument). The receivers (investors in the prior example or the news followers in the latter example) can rationally adopt the firm’s (or the news provider's) messages only if doing so does not worsen their state estimate relative to their prior knowledge, i.e., they follow messages only when an incentive-compatibility (IC) condition holds. On the other hand, the firm's (or the news agency's) goal is to use \emph{timeliness} of the information to maximize the fraction of time that the audience believes fundamentals (or the campaign, in the latter example) is doing well, subject to the budget \(R\) and the IC constraint. Motivated by these examples, the key research question that we aim to investigate in this work is: 

\textit{``By controlling only the timings of the information provided to the receiver, can the information provider (the sender) persuade the receiver to act in a way that the provider's utility is maximized?"}

In this environment, \emph{timeliness becomes the persuasion tool}: by accelerating good-news releases and delaying bad-news releases without altering truth, sender shifts real-time posteriors toward the favorable state while remaining within truthful–disclosure constraints.
Our setting is a Bayesian–persuasion problem where a sender shapes a receiver’s belief to influence actions. The classic foundations---Strategic Information Transmission (SIT) \cite{crawford1982strategic} and later as Bayesian persuasion \cite{kamenica2011bayesian}---study \emph{static} designs that specify \emph{which information} to reveal in a one-shot environment. By contrast, our persuasion lever is \emph{timeliness}: the sender chooses \emph{when to release} information by allocating state-dependent update rates over a CTMC. The rational receiver follows messages only when doing so does not worsen their estimate relative to prior information.

In dynamic approaches, \citet{ely2017beeps} shows how a sender \emph{schedules} disclosures to shape actions over evolving states. Subsequent work by \citet{che2023dynamic} studies frictions and timing costs, yielding Markov-perfect outcomes and links to static benchmarks. With commitment on a Markov state, optimal policies exhibit greedy/threshold forms and connect to sequential disclosure frameworks in \citet{renault2017optimal,au2015dynamic}. In more applied settings, dynamic persuasion has been studied with exogenous signals shaping timing incentives \citep{bizzotto2021outside}, receiver search and inspection generating persuasion-acquisition feedback \citep{yao2023dynamicsearch}, partial sender knowledge motivating “starting rough’’ \citep{nuta2024starting}, quadratic state-dependent costs in Gaussian models \citep{sayin2022quadratic}, and continuous-time filtering/control approaches that capture belief dynamics \citep{aid2025filtering}. Related notions of “timeliness’’ arise in models of optimal waiting \citep{orlov2016wait} and interim disclosure between mandatory announcement times \citep{gietzmann2023silence}. \emph{Closest to our setting}, \citet{ely2017beeps} treats timing as the instrument for what to disclose over time; \citet{ashkenazi2023markovtwo} study a two-state Markov environment with a myopic receiver and characterize intertemporal disclosure/silence rules; \citet{farhadi2022dynamic} develop dynamic information design over controlled Markov processes with planner-style signals and incentive constraints; and \citet{lehrer2025markov} analyze Markovian persuasion with stochastic revelations.

This rate-based formulation can also model real-time systems in which the timing of information plays a critical role. Recently, age of information (AoI) has been introduced to measure the timeliness of information in communication systems \citet{Kaul2011, Yates12}. Timely remote estimation problem for a Wiener process under a total sampling constraint have been considered in a seminal work of \cite{Sun2020}. The timely tracking of Poisson counting processes and infection status with exponential time intervals have been studied in \cite{Bastopcu_google, bastopcu2022using}. Recently, information sources have been modeled as Markov chains, and remote estimation problems have been studied to minimize the age of incorrect information (AoII) metric in \cite{maatouk2020age, cosandal2024multi, Pappas2025, luo2025roleagesemanticsinformation, saurav2025monitoring}. More specifically, \citet{LuoRemote_estimation} study the minimization of false and missed alarms in remote estimation of a binary Markov source. \citet{ayan2025ageawarecsiacquisitionfinitestate} examine the timely sharing of channel-state information (CSI) to jointly optimize communication performance and the cost of CSI acquisition. Timely task processing under state-dependent worker performance has been analyzed in the context of task completion efficiency in \cite{sariisik2025maximizeefficiencysystemsexhausted, liyanaarachchi2025ageestimatessubmitjobs}. Unlike timely remote estimation problems in the AoI literature, in our work, the sender and receiver have misaligned objectives; consequently, by strategically adjusting the timeliness of updates, the sender seeks to persuade the receiver to maintain its estimate in the desired state.

In this work, we first introduce our system model and formulate the persuasion problem over a 2-state CTMC in Section \ref{sect:system}. We then focus on the single-source setting and characterize the Stackelberg equilibrium in Section \ref{sect:solution1}. There, we show that the sender must transmit state-0 information at least at rate $c_{i,\min}$ to satisfy the IC constraint and can allocate the remaining rate budget to state-1 updates. In Section \ref{sect:multisourceproblem}, we extend our analysis to the multi-source setting, where each source may require a different $c_{i,\min}$ to satisfy its IC constraint. Consequently, the sender must decide which subset of sources to update and how to allocate its limited rate budget among them. We demonstrate that, for any fixed set of sources, the sender’s optimal rate allocation for state-1 updates is convex. Finally, in Section \ref{Num_result}, we provide extensive simulations showing that the sender can achieve persuasion solely through controlling the timeliness of information.

\begin{figure}[tb]
  \centering
  \includegraphics[width=0.93\columnwidth]{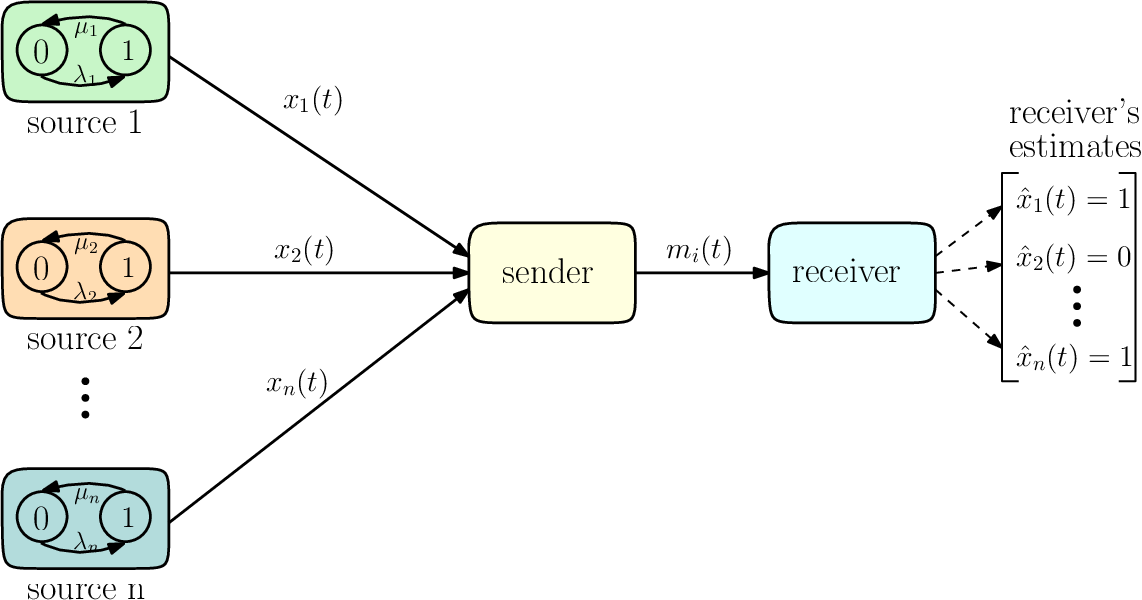}\vspace{-0.25cm}
  \caption{Communication system with $n$ sources, a sender, and a receiver.}
  \label{fig:system}
  \vspace{-0.15cm}
\end{figure}

\section{\!\!System Model and Problem Formulation}\label{sect:system}
\vspace{-0.1cm}
In this work, we consider a system composed of $n\geq 1$ information sources, a sender, and a receiver. Here, each source denoted by $I_i$ for $i = 1,\dots,n $  generates binary information streams (0's or 1's) with time-varying dynamics. More specifically, the binary information at source $I_i$ follows a 2-state CTMC where the transition from state 0 to state 1 happens with rate $\lambda_i\!>\! 0$ and from state 1 to state 0 with rate $\mu_i\! >\!0$ as shown in Fig.~\ref{fig:system}. We denote source $i$'s state at time $t$ as $x_i(t)$. Both the sender and the receiver know $(\lambda_i, \mu_i)$ for all $i$, but only the sender is capable of continuously observing these sources and share them with the receiver. On the other hand, since the receiver does not observe $x_i(t)$, with its prior knowledge, the receiver will only know the steady-state distribution of the CTMC which can be easily found by
\begin{align}
\pi^i_{0} = \frac{\mu_i}{\mu_i + \lambda_i},\qquad
\pi^i_{1} = \frac{\lambda_i}{\mu_i + \lambda_i}.
\label{eq:steady_state_probs}
\end{align}

The receiver would like to estimate the sources' state as accurately as possible. Based on the prior knowledge on the sources' rates and the information obtained from the sender, the receiver forms the real-time estimate about source $i$'s state $\hat{x}_i(t)$ at time $t$. With the sources' states and its corresponding estimates, the receiver will get the following utility $u_i(t) \!= \!q  \mathbbm{1} (x_i(t)\!\! =\! 0, \hat{x}_i(t) \!\!=\! 0) \!+\!(1\!-\!q)  \mathbbm{1}(x_i(t)\!$ $ =\! 1, \hat{x}_i(t)\! = \!1) $ from source $i$ as shown in Table~\ref{Table:rec_utility} where $\mathbbm{1}(.)$ is the indicator function returning 1 when $(.)$ is true. 
\begin{table}
   \caption{The receiver's utility function $u_i(t)$.}
    \label{Table:rec_utility}
    \centering
    \begin{tabular}{c|c|c}
        $(x_i(t), \hat{x}_i(t))$ & 0     & 1     \\ \hline
        0                        & $q$   & 0     \\ \hline
        1                        & 0     & $1-q$ \\
    \end{tabular}
    \vspace{-0.2cm}
\end{table}  

In other words, when $x_i(t) = 0,$ $ \hat{x}_i(t) = 0$, the receiver will obtain a weighted reward, that is, $u_i(t) = q$ and when $x_i(t) = 1,$ $ \hat{x}_i(t) = 1$, it will receive $u_i(t) =1- q$ from source $i$ where $0<q<1$. When there is no information provided, the receiver will only know the steady state distribution of $x_i(t)$, i.e., $(\pi^i_{0},\pi^i_{1})$ for all $i$ in (\ref{eq:steady_state_probs}). We assume that $ q\pi^{i}_{0} = \frac{q\mu_i}{\mu_i + \lambda_i} > \frac{(1-q)\lambda_i}{\mu_i + \lambda_i} =(1-q) \pi^i_{1}$ for all $i$. As a result, when there is no information provided, we assume that the receiver's default estimation is $\hat{x}_i(t) = 0$ for all $t$. Similarly, the sender's utility obtained from source $i$ at time $t$ is denoted as $v_i(t)$ and is given in Table~\ref{Table:sender_utility}. As opposed to the receiver, the sender's utility will be equal to 1 only when the receiver's estimate is equal to $\hat{x}_i(t) = 1$ irrespective of the source $i$'s state.  
\begin{table}
   \centering
    \caption{The sender's utility function $v_i(t).$}\label{Table:sender_utility}

    \begin{tabular}{c|c|c}
        $(x_i(t), \hat{x}_i(t))$ & 0     & 1     \\ \hline
        0                        & 0   & 1     \\ \hline
        1                        & 0     & 1 \\
    \end{tabular}
\vspace{-0.2cm}
\end{table}

Different from the most traditional communication literature where the sender and the receiver have aligned goals, here, we note from Tables~\ref{Table:rec_utility} and \ref{Table:sender_utility} that the sender's and the receiver's utility functions are different and the sender's utility depends also on the receiver's estimate at time $t$. More specifically, while the receiver wants to know the states as accurately as possible, the sender wants the receiver to always estimate the state as $\hat{x}_i(t) =1$ for all $t$. Different from the SIT and Bayesian persuasion literature where the sender can  modify the source's correct information, in this work, we consider persuasion via information timeliness. That is, when source $i$’s state flips between 0 and 1, the sender dynamically modifies its transmission rate by accelerating or decelerating communication to steer the receiver’s estimation of the current state.

To model such a system, we consider a setting where the sender shares the sources' states with the receiver with random time intervals. More specifically, when the source $i$'s state is equal to 0, we model the sender’s inter-transmission times as exponentially distributed with rate $c_i\geq 0$. Similarly, when the source $i$'s state is equal to 1, we model the sender’s inter-transmission times as exponentially distributed with rate $s_i\geq 0$. Let us denote the time instance where the sender sends the $j$th update (where $j\geq1$) about the $i$th source's  state as $t_{ij}$. By denoting the sender's $j$th message about source $i$'s state as $m_{ij} = x_i(t_{ij})$, under the condition that the receiver follows the sender's messages (which we will specify this condition precisely), the receiver will form the following estimate $\hat{x}_i(t)$ at time $t$ based on the received messages: 
\begin{align}\label{est_process}
    \hat{x}_i(t) = m_{ij}, \qquad t_{i,j}\leq t<t_{i,j+1}, 
\end{align}
where we assume that the receiver knows the initial values of the sources, i.e., $m_{i0} = x_i(0)$ and $t_{i,0} = 0$ for all $i$.

As the objectives of the sender and the receiver are different, we formulate the interaction between these agents as a Stackelberg game, in which the sender acts as the leader and the receiver as the follower. In this Stackelberg game, the sender commits to a strategy first by choosing $\mathbf{\beta} = \{\beta_1, \cdots, \beta_n\}$ where $\beta_i = (s_i,c_i)$ for all $i$.\footnote{In this work, we restrict the sender’s policy space to Poisson sampling and characterize the corresponding Stackelberg equilibrium. Focusing on Poisson sampling policies allows for analytical tractability and has been considered in the literature, such as \cite{Bastopcu_google, Akar_CTMC}. } Then, the receiver observes the sender's information revealing policy and selects its best response. At this point, the receiver has two options: ({\it{i}}) if the sender's messages about source $i$ leads to an estimate no worse than the initial knowledge to maximize the receiver's long-term average utility, i.e.,  $\lim_{T\rightarrow\infty}\!\!\frac{1}{T}\!\int_{t=0}^{T}\! u_i(t) dt $, the receiver will follow the sender's messages as in (\ref{est_process}). We denote this policy as $\sigma_{\text{sender}}$. ({\it{ii}}) If following the sender's messages leads to a lower average utility compared to the default policy which uses only the prior information, then the receiver will ignore the sender's messages for source $i$ and use the estimate $\hat{x}_i(t) = 0$ for all $t$ which will give the utility of $ q\pi^{i}_{0} = \frac{q\mu_i}{\mu_i + \lambda_i}$ from source $i$. We denote this policy as $\sigma_{\text{default}}$.

Finally, we represent the sender's utility function obtained from source $i$ as $J_{S,i}(\beta_i,  BR(\beta_i))= \lim_{T\rightarrow\infty}\frac{1}{T}\int_{t=0}^{T}v_i(t) dt$ and the sender's total average utility as $J_S(\mathbf{\beta},  BR(\mathbf{\beta})) = \sum_{i=1}^n J_{S,i}(\beta_i,  BR(\beta_i))$ which depends on the sender's committed policy $\mathbf{\beta}$ and the receiver's best response to $\mathbf{\beta}$ given by $BR_i(\mathbf{\beta}_i)\in\{\sigma_{\text{sender}},$ $ \sigma_{\text{default}}\}$. Similarly, based on the sender's policy, if the receiver follows the sender's messages as in (\ref{est_process}), the receiver will obtain the average utility of $J_{R,i}(\beta_i)=\lim_{T\rightarrow\infty}\frac{1}{T}\int_{t=0}^{T}u_i(t) dt $ from source $i$ where the receiver's estimate $\hat{x}_{i}(t)$ is determined in (\ref{est_process}) based on the policy $\beta_i$. Thus, we define the Stackelberg equilibrium as
\begin{align}\label{eqn:Stackelberg}
    J_S(\mathbf{\beta}^*,  BR(\mathbf{\beta}^*) ) \geq & J_S(\mathbf{\beta},  BR(\mathbf{\beta}) )
    \end{align}
where
\begin{align}\label{eqn:Stackelberg_2}
BR_i(\beta_i)=
\begin{cases}
\sigma_{\text{sender}}, &
J_{R,i}(\beta_i)\geq
\dfrac{q\mu_i}{\mu_i+\lambda_i},\\
\sigma_{\text{default}}, &
J_{R,i}(\beta_i)<
\dfrac{q\mu_i}{\mu_i+\lambda_i}.
\end{cases}
\end{align}
In other words, the Stackelberg equilibrium in (\ref{eqn:Stackelberg}) and (\ref{eqn:Stackelberg_2}) is achieved when the sender commits to a policy that will maximize its own utility function by  also considering how the receiver will respond to the sender's committed policy. Next, based on the sender's policy $\beta_i= (s_i,c_i)$, we first   characterize the receiver's cost function $J_{R,i}(\beta_i)$. 
\begin{figure}[tb]
  \centering
  \includegraphics[width=0.385\columnwidth]{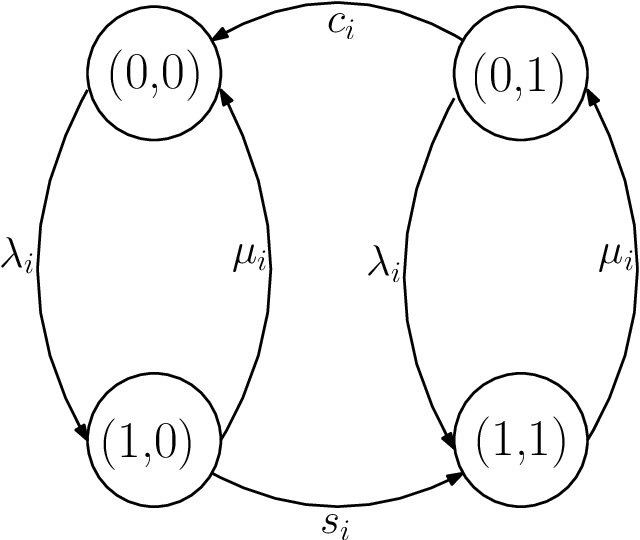}
  \vspace{-0.40cm}
  \caption{Continuous-time Markov chain for $(x_i(t), \hat{x}_i(t))$.}
  \label{fig:CTMC}
  \vspace{-0.10cm}
\end{figure}
  \vspace{-0.2cm}
\subsection{The Receiver's Average Utility Function}
  \vspace{-0.2cm}
We note that as a result of following the sender's messages, source $i$'s state and its estimate at the receiver $(x_i(t), \hat{x}_i(t))$ form a CTMC with four states given as $\{(0,0),(0,1),(1,0), (1,1)\}$ as shown in Fig.~\ref{fig:CTMC}. By dropping source $i$'s index to derive a general expression, similar to the steps in \cite{bastopcu2022using}, we find the unique stationary distribution of the CTMC given by $\pi = \{\pi_{00},\ \pi_{01},\ \pi_{10},\ \pi_{11}\}$. For that, we first write the local-balance equations as:
\begin{align} \label{eq:LocalBalanceEq}
   \pi_{00} \lambda &= \pi_{10} \mu + \pi_{01} c, \\
     \pi_{10}( \mu+  s) &= \pi_{00} \lambda, \\
     \pi_{01} (c +\lambda)  &= \pi_{11} \mu,
     \end{align}
 \begin{align}     
     \pi_{11} \mu &= \pi_{10} s + \pi_{01} \lambda .
\end{align}
Using the above equations and $ 
\sum_{m=0}^{1} \sum_{n=0}^{1}\! \pi_{mn} = 1, 
$ we find the steady-state distribution of the CTMC as:
\begin{align*}
\pi_{00} \!= \!\frac{\mu c (\mu\!+\!s)}{\kappa},~ \pi_{01}\! =\! \frac{\mu  \lambda s}{\kappa},~ \pi_{10} \!= \!\frac{\mu  \lambda c}{\kappa},~ \pi_{11}\! =\! \frac{\lambda s (\lambda \!+\!c) }{\kappa},
\end{align*}
where $\kappa = (\mu + \lambda)(\mu c + \lambda s + c s) $. Then, from Table~\ref{Table:rec_utility}, we can find the long-term average of the receiver's utility obtained from source $i$ as
\begin{align}
J_{R,i}(\beta_i)
&= q\,\pi^{i}_{00} + (1-q)\,\pi^{i}_{11} \notag\\
&= \frac{ q\,\mu_i c_i (\mu_i+s_i) + (1-q)\,\lambda_i s_i (\lambda_i + c_i)}
         { (\mu_i+\lambda_i)\bigl(\mu_i c_i + \lambda_i s_i + c_i s_i\bigr)}.
\label{eqn:rec_cost_closed_form}
\end{align}

Note that
if the messages $m_{ij}$ help the receiver to form a better estimate (which will be equivalent to the incentive compatibility (IC) constraint in Bayesian Persuasion \cite{kamenica2011bayesian}), the receiver will follow the messages $m_{ij}$ as in (\ref{est_process}) and as a result obtain $J_{R,i}(\beta_i)$ in (\ref{eqn:rec_cost_closed_form}). Otherwise, the receiver will use $\hat{x}_i(t) = 0$ for all $t$ and obtain the utility of $\frac{q\mu_i}{\mu_i + \lambda_i}$ from source $i$.  
\vspace{-0.2cm}
\subsection{Sender's Persuasion Problem}
\vspace{-0.2cm} 
As mentioned earlier, the sender's utility function
$v_i(t)$ provided in Table~\ref{Table:sender_utility} depends on the receiver's estimation. As a result, the sender wants to influence the receiver to follow its messages and affect $\hat{x}_i(t)$ in a way to maximize its own utility. To do that, the sender should commit to a policy  $\beta_i = (s_i,c_i)$ such that the receiver's utility $J_{R,i}(\beta_i)$ is greater than or equal to $ \frac{q\mu_i}{\mu_i + \lambda_i}$ which is the IC constraint. When we substitute $J_{R,i}(\beta_i)$ provided in (\ref{eqn:rec_cost_closed_form}) into $J_{R,i}(\beta_i)\geq \frac{q\mu_i}{\mu_i + \lambda_i}$ and after some algebraic manipulations, under the assumption that $s_i>0$, we have 
\begin{align}\label{eqn:IC_constraint}
    c_i\geq c_{i,\min} =  \frac{q\mu_i}{1-q}-\lambda_i. 
\end{align}
We note that due to our assumption that the receiver's initial estimation without sender's information is equal to 0, that is $ q\pi^{i}_{0} = \frac{q\mu_i}{\mu_i + \lambda_i} > \frac{(1-q)\lambda_i}{\mu_i + \lambda_i} =(1-q) \pi^i_{1}$, $c_{i,\min}$ will always be positive, i.e., $c_{i,\min}>0$. On the other hand, when $s_i = 0$, this constraint is automatically satisfied for all $c_i$. Then, the sender's persuasion problem becomes: 
\begin{align}
\label{eqn:senders_problem}
\max_{\{ s_i, c_i \}}  \quad & \sum_{i=1}^{n}(\pi^i_{01} + \pi^i_{11})  \!\!= \!\!\sum_{i=1}^{n}\frac{\lambda_i s_i (c_i + \lambda_i + \mu_i)}{(\mu_i + \lambda_i)(\mu_i c_i + \lambda_i s_i + c_i s_i)} \nonumber \\
\mbox{s.t.} \quad & \sum_{i=1}^{n} (c_i + s_i) \leq R \nonumber \\
\quad &  c_i\geq \mathbbm{1}(s_i>0) c_{i,\min}, ~ s_i \geq 0, ~~ i \in\{1,\ldots,n\}.\!\!\!
\end{align}

We note from (\ref{eqn:senders_problem}) that the sender has a total sampling constraint $R$ such that $\!\sum_{i\!=\!1}^{n}\!\! c_i +\! s_i\!\! \leq\! \!R$. The second constraint ($c_i\!\geq\! \mathbbm{1}(s_i\!>\!0) c_{i,\min}$) in  (\ref{eqn:senders_problem}) is the IC constraint for each source and the third constraint is the feasibility constraint.  

In the next section, we provide the Stackelberg equilibrium of the game formulated in (\ref{eqn:Stackelberg}) and (\ref{eqn:Stackelberg_2}) for a single source. 

\vspace{-0.15cm}
\section{The Optimal Information Revelation Policy for a Single-Source Case}\label{sect:solution1}
\vspace{-0.15cm}

In this section, we provide an explicit solution to the sender's information revelation problem in (\ref{eqn:senders_problem}) and thereby characterize the Stackelberg equilibrium of the game formulated in (\ref{eqn:Stackelberg}) and (\ref{eqn:Stackelberg_2}). In order to understand the sender's policy better,  we first focus our attention on the single source problem. For convenience, by dropping the source index $i$, we rewrite the sender's optimization problem for a single source, i.e., the case with $n=1$, as
\begin{align}
\label{eqn:senders_problem_n_1}
\max_{\{ s, c \}}  \quad & \pi_{01} + \pi_{11}  = \frac{\lambda s (c + \lambda + \mu)}{(\mu + \lambda)(\mu c + \lambda s + c s)} \nonumber \\
\mbox{s.t.} \quad & c + s \leq R \nonumber \\
\quad &  c\geq \mathbbm{1}(s>0) c_{\min}, \quad s \geq 0.
\end{align}

As seen in (\ref{eqn:senders_problem_n_1}), there is a certain minimum $c$ sampling rate denoted as $c_{\min} = \frac{q \mu}{1-q}-\lambda$ coming from the IC condition which is strictly positive as mentioned before. In order to find the sender's optimal solution, in the next lemma, we characterize the sender's utility function's behavior with respect to the sampling rates $s$ and $c$.
\begin{lemma}\label{Lemma_1}
 Under the assumption that the IC condition holds, the sender's utility function $J_S(s,c) \!\!= \!\!\frac{\lambda s (c + \lambda + \mu)}{(\mu + \lambda)(\mu c + \lambda s + c s)}$   is an increasing function of $s$ and a decreasing function of $c$ when $s>0$.   
\end{lemma} 
\begin{Proof}
We begin the proof by showing that $J_S(s,c)$ is an increasing function of $s$ when $c>0$. For that, the partial derivative of $J_S(s,c)$ with respect to $s$ is given by 
\begin{align*}
    \frac{\partial J_S(s,c)}{\partial s} = \frac{\lambda\mu c( \lambda + \mu+c)}{(\mu + \lambda)(\mu c + \lambda s + c s)^2}.
\end{align*}
Thus, we have $\frac{\partial J_S(s,c)}{\partial s}>0$ when $c>0$ which is indeed the case due to the IC constraint which implies that $c\geq c_{\min}>0$. As a result, $J_S(s,c)$ is an increasing function of $s$. Similarly, to show that $J_S(s,c)$ is a decreasing function of $c$, we find $\frac{\partial J_S(s,c)}{\partial c}$ as
\begin{align*}
    \frac{\partial J_S(s,c)}{\partial c} = -\frac{\lambda \mu s(\lambda+\mu+s)}{(\mu+\lambda)(\mu c+\lambda s + cs)^2}.
\end{align*}
As a result, we have $ \frac{\partial J_S(s,c)}{\partial c} <0$ when  $s>0$ which completes the proof. 
\end{Proof}

Thus, if possible, to maximize the sender's utility, the sender should allocate most of its  sampling rate to $s$ and some of its sampling rate to $c$ to meet the IC constraint. In the next theorem, we completely characterize the Stackelberg equilibrium of the game in the case of a single source.
\begin{theorem}
    The Stackelberg equilibrium of the single-source game is given by 
    \begin{align*}
        (\beta^*, BR(\beta^*))  =\begin{cases} 
      ((0,0), \sigma_{\text{default}}),  & R< c_{\min} \\
      ((R-c_{\min},c_{\min}), \sigma_{\text{sender}}),& R\geq  c_{\min}
   \end{cases} 
    \end{align*}
As a result of this Stackelberg equilibrium, the sender will  obtain the utility of: 
\begin{align*}
        J_S(\beta^*\!\!,\! BR(\beta^*))  \!\!=\!\!\begin{cases} 
      0,  &\!\!\!\text{if } R\!<\! c_{\min} \\
      \!\!\frac{\lambda \bar{s} (c_{\min}+\lambda+\mu)}{(\mu\!+\!\lambda)(\mu c_{\min}\!\! +\! \lambda \bar{s} \!+\! c_{\min} \bar{s})},&\!\!\!\text{if } R\!\geq \! c_{\min}
   \end{cases}  
    \end{align*}
where $\bar{s} = R -c_{\min}$. The receiver will obtain the utility of $J_R(\beta^*, BR(\beta^*)) = \frac{q\mu}{\mu+\lambda}$.
\end{theorem}
\begin{Proof}
We begin our proof by considering the case when $R\!<\!c_{\min}$. In this case, the sender cannot pass the minimum required sampling rate $c_{\min}$ for the receiver to follow the sender's messages. As a result, for all the policies $\beta$ that the sender can commit, the receiver would choose $BR(\beta) = \sigma_{\text{default}}$. In other words, the sender cannot meet the IC constraint and as a result, the receiver would not follow the sender's messages and keep its estimate $\hat{x}(t) = 0$ for all $t$ in which case the sender would get 0 utility. For the sender, since all the policies would yield the same utility, we choose $\beta^* \!=\! (s^*,c^*) \!=\! (0,0)$ as the sender's policy when $R<c_{\min}$. As a result of this equilibrium, the sender obtains $J_S(\beta^*, BR(\beta^*)) = 0$ and the receiver obtains $J_R((0,0),\sigma_{\text{default}}) = \frac{q\mu}{\mu+\lambda}$ utilities, respectively.

Next, we consider the case when $R\!\geq\! c_{\min}. $ In this case, the sender has sufficient total sampling rate such that it can persuade the receiver to follow its messages. Due to Lemma~\ref{Lemma_1}, the sender's utility is an increasing function of $s$ and a decreasing function of $c$ when $s>0$ and $c>0$. As a result, the sender should allocate $c\!=\! c_{\min}$ to satisfy the IC constraint, then allocate the remaining sampling rate to $s$ to maximize its own utility. Hence, when $R\geq c_{\min}$, the Stackelberg equilibrium is achieved at $(\beta^*, BR(\beta^*))=((R-c_{\min},c_{\min}), \sigma_{\text{sender}})$. The corresponding utilities of the sender and the receiver are given by $J_S(\beta^*, BR(\beta^*)) = \frac{\lambda \bar{s} (c_{\min}+\lambda+\mu)}{(\mu+\lambda)(\mu c_{\min} + \lambda \bar{s} + c_{\min} \bar{s})}$ with $\bar{s} = R-c_{\min}$ and $J_R(\beta^*, BR(\beta^*)) =\frac{q\mu}{\mu + \lambda}$, respectively. 
\end{Proof}

Therefore, for a single-source Stackelberg game, if the sender has sufficiently large sampling rate $R$, it should allocate the minimum sampling rate $c=c_{\min}$ to send the source's state-0 information to provide sufficient information to the receiver for persuasion. Then, it should allocate its remaining sampling rate $s= R-c_{\min}$ for sampling the source's state-1 information to increase its own utility. As a result of applying this policy, the receiver follows the sender's messages, i.e., $\sigma_{\text{sender}}$, and the receiver's utility will be at the same level compared to the case only with the prior information.\footnote{When $R\!\!\geq \!\!c_{\min}$, the receiver obtains the same utility under both policies, $\sigma_{\text{sender}}$ and $\sigma_{\text{default}}$. Throughout this work, we adopt an optimistic assumption that, when the receiver faces multiple actions yielding equal utility, it chooses the action that benefits the sender the most. Under a pessimistic approach, the sender would set $c \!=\! c_{\min}\!+\!\epsilon$ and $s\! \!=\!\! R\!\!-\!\! (c_{\min}\!\!+\!\epsilon)$ for an arbitrarily small $\epsilon\!\!>\!\!0$, ensuring that $\sigma_{\text{sender}}$ provides a slightly higher utility. Yet, as $\epsilon\!\!\rightarrow \!0$, the receiver’s utility becomes the same value as in the optimistic one.} On the other hand, the sender benefits from applying this policy as  the receiver's estimate is equal to $1$ for some portion of the time. 

Building on the insights from this section, we next generalize the results from the single-source case to the multi-source setting.

\vspace{-0.15cm}
\section{The Optimal Information Revelation Policy for a Multi-Source Case}\label{sect:multisourceproblem}
\vspace{-0.15cm}
In this section, we extend our analysis to a setting where the sender reveals information about multiple sources to the receiver. Our goal is to solve the general persuasion problem with $n\geq 1$ sources in (\ref{eqn:senders_problem}). The sender's information revelation policy is more involved since the sender should decide which information source it should sample and at which rates. In order to characterize the sender's optimal information revelation policy, we start with the case when the sender's total sampling rate is limited such that we have $R<\min_{i\in\{1,\cdots,n\}} c_{i,\min}$. 
\begin{lemma}\label{Lemma2}
    When $R<\min_{i\in\{1,\cdots,n\}} c_{i,\min}$, the Stackelberg equilibrium of the game is achieved when $\beta_i^* = (s_i^*,c_i^*)=(0,0)$  and $ BR_i(\beta_i^*) = \sigma_{\text{default}}$ for all $i$. In this case, the sender obtains $J_S(\beta^*, BR(\beta^*)) = 0$ and the receiver obtains $J_R(\sigma_{\text{default}}) = \sum_{i=1}^{n}\frac{q\mu_i}{\mu_i+\lambda_i}$.  
\end{lemma}
\begin{Proof}
    The proof of Lemma~\ref{Lemma2} directly follows from the IC constraint. In this case, since the total sampling rate of the sender is limited such that $R<\min_{i\in\{1,\cdots,n\}} c_{i,\min}$, the sender cannot allocate sufficient rate for any source to persuade the receiver to follow its messages. Thus, the receiver chooses $ BR_i(\beta_i) = \sigma_{\text{default}}$ for all $i$ for any policy of the sender $\beta_i$. Although all the sender's policies  would give the same receiver's best response,  that is, $ BR_i(\beta_i) = \sigma_{\text{default}}$ for all $i$, we particularly choose the sender's policy to be $\beta_i^* = (s_i^*,c_i^*)=(0,0)$ for convenience. As a result, the Stackelberg equilibrium is achieved when $\beta_i^* = (s_i^*,c_i^*)=(0,0)$ and $ BR_i(\beta_i^*) = \sigma_{\text{default}}$ for all $i$.   
\end{Proof}

In the remaining part of this subsection, we focus our attention on the case when $\!R\!>\!\! \min_{i\in\{1,\cdots,n\}} \!c_{i,\min}$.\footnote{ Note that when $ \min_{i\in\{1,\cdots,n\}} c_{i,\min}\!\!=\!\!R$, the sender can meet the IC constraint for some sources, but since there is no sampling rate remains to allocate to $s_i$, the sender still gets 0 utility in this case.} Thus, at least by allocating all of its sampling rate, the sender is capable of persuading the receiver to follow its messages for some sources. To characterize the sender's optimal information revelation policy, next, we state that the sender decides to send information about source $i$'s 1-state with the sampling rate $s_i\!>\!0$ if and only if the sampling rate for source $i$'s 0-state $c_i$ must be equal to   $c_{i,\min}$.   
\begin{lemma}\label{Lemma3}
    When $R\!>\!\! \min_{i\in\{1,\cdots,n\}} c_{i,\min}$, we have $s_i\!>\!0$ if and only if $c_i = c_{i,\min}$ for the sender's optimal policy.  
\end{lemma}
\begin{Proof}
    The proof of Lemma~\ref{Lemma3} is an immediate consequence of the IC constraint and Lemma~\ref{Lemma_1}. Assume by contradiction that for the sender's optimal sampling policy, there exist sampling rates with $s_i>0$ and $c_i < c_{i,\min}$ for some $i$. For such sources, the receiver does not follow the sender's messages since the IC constraint is not met. As a result, the sender would obtain $J_{S,i}(\beta_i) = 0$ from these sources. Then, let us consider all the sources $j$ with $c_{j,\min}< R $ and $\mathcal{J}$ is the index set of all such sources, that is, $j\in \mathcal{J}$. In the existing policy, if there is already source $j$ with $c_j= c_{j,\min}$, then we can obtain a strictly higher utility by choosing $s_i = 0$ and $c_i = 0$ for all the sources with $s_i>0$ and $c_i < c_{i,\min}$ and allocate these sampling rates to source $j$ to increase $s_j$.  In the existing policy, if there is no such source with $c_j= c_{j,\min}$, then it means that the sender's utility is equal to 0 since the IC constraint is not met for any sources. In this case again, by choosing $s_i = 0$ and $c_i = 0$ for all the sources with $s_i>0$ and $c_i < c_{i,\min}$, the sender can allocate these sampling rates to a source in $j\in \mathcal{J}$ to make $c_j = c_{j,\min}$ and allocate remaining sampling rates to $s_j$. Since in both cases, the new proposed policy gives a strictly higher utility, we reach a contradiction. Thus, if $s_i>0$, then we must have $c_i = c_{i,\min}$.\footnote{Note that Lemma~\ref{Lemma_1} implies that, under the sender's optimal policy, $c_i$ cannot exceed $c_{i,\min}$ whenever $s_i>0$.}
    
    Similarly, if $c_i = c_{i,\min}$, then we must have $s_i>0$. For the sources with $c_i = c_{i,\min}$ and $s_i=0$, the sender obtains $J_{S,i}(\beta_i) = 0$ from these sources. Following a similar argument as above, one can readily show that the sender can achieve a strictly higher utility by reallocating its sampling rates to the sources $j\in \mathcal{J}$. Therefore, when $R> \min_{i\in\{1,\cdots,n\}} c_{i,\min}$, for the sender's optimal sampling policy, we have $s_i\!>\!0$ if and only if $c_i = c_{i,\min}$.            
\end{Proof}
Thus, Lemma~\ref{Lemma3} states that if the sender decides to allocate non-zero sampling rates, then the sampling rates must be $c_i =c_{i,\min}$ (to satisfy the IC constraint) and $s_i> 0$. Based on these initial results, next we will re-formulate the sender's persuasion problem in (\ref{eqn:senders_problem}) as follows: 
\begin{align}
\label{eqn:senders_problem_mod}
\max_{\{ x_i, s_i \}}  \quad & \hat{J_S}(\mathbf{x},\mathbf{s}) = \sum_{i=1}^{n}  x_i\frac{\lambda_i}{\lambda_i + \mu_i}  \frac{s_i (c_{i, \min} + \mu_i + \lambda_i)}{c_{i, \min} \mu_i + s_i \lambda_i + c_{i, \min} s_i} \nonumber \\
\mbox{s.t.} \quad & 
\sum_{i=1}^{n} \left( x_i c_{i, \min} + s_i \right) \leq R\nonumber \\
\quad &  x_i \in \{0, 1\}, ~~~s_i \geq 0,~~~ i \in\{1,\ldots,n\}.
\end{align}

In the reformulation, we exploit the fact that $c_i$ can take only two values: $c_{i,\min}$ or $0$. To model this binary behavior, we introduce a decision variable $x_i \!\!\in \!\!\{0,1\}$ in (\ref{eqn:senders_problem_mod}), where $x_i\! = \!1$ corresponds to $c_i\!\! =\!\! c_{i,\min}$, and $x_i \!=\! 0$ corresponds to $c_i\! =\! 0$. When $x_i \!\!=\!\! 0$ (i.e., $c_i\!\!=\!\! 0$),  the sender receives zero utility from that source, as the IC constraint is not satisfied. Therefore, multiplying the objective function by $x_i$ in (\ref{eqn:senders_problem_mod}) appropriately captures this behavior. Similarly, the total sampling constraint can be written as $\sum_{i=1}^{n} \!\!\left( x_i c_{i, \min} \!+\! s_i \right) \!\leq\! R$. With this reformulation, it is also easy to see that if $x_i = 0$, then we have $s_i\!=\! 0$ as the objective function for that source will be equal to zero due to $x_i\!=\!0$ multiplier in the objective.

To solve the optimization in (\ref{eqn:senders_problem_mod}), we fix the values of $x_i$ and then solve the resulting optimization problem over the parameters $s_i$. With this goal, for a given set of $x_i$'s, first, we analyze the convexity of the  sender's persuasion problem in (\ref{eqn:senders_problem_mod}) with respect to $s_i$.

\begin{lemma}\label{Lemma4}
For a given set of $x_i$'s,  the sender's information revelation problem in (\ref{eqn:senders_problem_mod}) is a convex optimization problem with respect to $s_i$.
\end{lemma}
\begin{Proof}
The first derivative of the sender's utility function is given by 
\begin{align*}
    \frac{\partial \hat{J}(\mathbf{x},\mathbf{s})}{\partial s_i}= x_i \frac{\lambda_ic_{i, \min} \mu_i}{\lambda_i +\mu_i} \frac{(c_{i, \min} + \mu_i + \lambda_i)}{(c_{i, \min} \mu_i + s_i\lambda_i + s_ic_{i, \min} )^2}.
\end{align*}
Similarly, the second derivative of the sender's utility function is given by
\begin{align*}
    \frac{\partial^2 \hat{J}(\mathbf{x},\mathbf{s})}{\partial s_i^2}= -2x_i \frac{\lambda_ic_{i, \min} \mu_i}{\lambda_i +\mu_i} \frac{(\lambda_i + c_{i, \min})(c_{i, \min} + \mu_i + \lambda_i)}{(c_{i, \min} \mu_i + s_i\lambda_i + s_ic_{i, \min} )^3}.
\end{align*}
Since the first derivative is non-negative and the second derivative is non-positive, that is, $\frac{\partial \hat{J}(\mathbf{x},\mathbf{s})}{\partial s_i}\geq 0$ and  $\frac{\partial^2 \hat{J}(\mathbf{x},\mathbf{s})}{\partial s_i^2}\leq 0$, respectively, we can conclude that the sender's utility is a concave non-decreasing function of $s_i$. Since the total sampling rate constraint, $\sum_{i=1}^{n} \left( x_i c_{i,\min} + s_i \right) \leq R$,\footnote{For some given sets of $x_i$'s, we may have $R- \sum_{i=1}^{n}x_i c_{i,\min} \leq 0$ in which case the problem can be infeasible.} and the feasibility constraint, $s_i \geq 0$, define a convex feasible region, the optimization problem in (\ref{eqn:senders_problem_mod}) is convex for any fixed set of $x_i$ values.    
\end{Proof}
For a given set of $x_i$'s, let us denote $\mathcal{S}$ as the set of source indices such that $x_i = 1$. Then, the complement of the set $\mathcal{S}$ given by $\mathcal{S}^c$ is the source indices with $x_i = 0$. We note that for the sources that are in the set $\mathcal{S}^c$, we have $s_i = 0$. For the remaining sources that are in the set $ \mathcal{S}$, in order to find their optimum update rates $s_i$, we introduce the Lagrangian function \cite{boyd2004convex}  for  (\ref{eqn:senders_problem_mod}) as follows:
\begin{align}
    \mathcal{L} =& -\sum_{i\in \mathcal{S}}  \frac{\lambda_i}{\lambda_i + \mu_i}  \frac{s_i (c_{i, \min} + \mu_i + \lambda_i)}{c_{i, \min} \mu_i + s_i \lambda_i + c_{i,\min} s_i}\nonumber\\& +\theta \left(\sum_{i\in \mathcal{S}} \left( c_{i, \min} + s_i \right)- R\right) - \sum_{i\in \mathcal{S}} \nu_i s_i,
\end{align}
where $\theta\geq 0$ and $\nu_i \geq 0$ for all $i$. Next, the KKT conditions are given by 
\begin{align}\label{eqn:KKT_cond}
\!\!\!\!\frac{\partial \mathcal{L}}{\partial s_i} \!\!=\!\! - \frac{\lambda_ic_{i, \min} \mu_i}{\lambda_i +\mu_i} \frac{(c_{i, \min} + \mu_i + \lambda_i)}{(c_{i, \min} \mu_i \!+ \!s_i\lambda_i \!+\! s_ic_{i, \min} )^2} \!+\! \theta\!-\!\nu_i  \!\!=\!\! 0,\!\!   
\end{align}
for all $i\in \mathcal{S}$. Then, the complementary slackness (C.S.) conditions can be stated as follows:  
\begin{align}
    \theta \left(\sum_{i\in \mathcal{S}} \left( c_{i, \min} + s_i \right)- R\right)  =& 0,\label{eqn:CS_1}\\
    \nu_i s_i =& 0,\label{eqn:CS_2}
\end{align}
for all $i\in \mathcal{S}$. By solving (\ref{eqn:KKT_cond}) for $s_i$, we obtain
\begin{align}\label{eqn:s_i_temp}
    s_i = C_i \left(\sqrt{\frac{A_i}{B_i(\theta- \nu_i)}} -1\right),
\end{align}
where $A_i = \lambda_i (c_{i, \min} + \mu_i + \lambda_i)$, $B_i = c_{i,\min}\mu_i (\lambda_i+\mu_i)$, and $C_i = \frac{c_{i,\min}\mu_i}{\lambda_i+c_{i,\min}}$. Due to the CS condition in (\ref{eqn:CS_2}), either $s_i > 0$ which implies $\nu_i = 0$ or $s_i = 0$ and we have $\nu_i \geq 0$. Thus, the optimal values of $s_i$ denoted by $s_i^*$ are equal to: 
\begin{align}\label{eqn:s_i_opt}
    s_i^* = C_i \left(\sqrt{\frac{A_i}{B_i\theta}} -1\right)^+,
\end{align}
where $(x)^+ = x$ if $x\geq 0$; $(x)^+ = 0$, otherwise. Although the closed-form solution for each \( s_i^* \) is known, it depends on \( \theta \), which must be chosen such that the total resource constraint \( \sum_{i \in \mathcal{S}} c_{i,\min}+s_i^* = R \) is satisfied. We note from (\ref{eqn:s_i_opt}) that  \( s_i^*\) is a strictly decreasing function of \( \theta \). As a result, the total allocation is also strictly decreasing in \( \theta \), making bisection search an appropriate method. We initialize the lower bound as \( \theta = 0 \) and the upper bound as \( \max_{i \in \mathcal{S}} \frac{A_i}{B_i} \), ensuring that only sources with \( \frac{A_i}{B_i} > \theta \) receive nonzero allocation. At each step, we compute the midpoint \( \theta \), evaluate the total allocation, and update the bounds depending on whether the sum is greater or less than \( R \). The process continues until convergence, yielding the unique \( \theta \) that balances total allocation exactly to the remaining budget. If the bisection method cannot find an optimal $\theta$ then we can conclude that this subset of active sources is not feasible, thus the subset should be changed.

From the expression of $s_i^*$ in (\ref{eqn:s_i_opt}), we observe that the optimal policy exhibits a threshold structure, i.e., not all update rates $s_i^*$ are necessarily positive. Depending on the values of $A_i$ and $B_i$, some sources may have an update rate of zero. If some sources have zero allocation, we can determine their order based on the ratio of $\frac{A_i}{B_i}$. Note that we find the optimal $s_i^*$ allocation for a given set of $x_i$'s. For a given set of $x_i$'s, if we allocate $s_i^* = 0$ for some of the sources that are in set $\mathcal{S}$, that is, $x_i=1$, such allocations cannot be globally optimal since the sender allocates $c_i = c_{i,\min}$ and $s_i=0$ for these sources. Due to Lemma~\ref{Lemma3}, $s_i=0$ if and only if $c_i =0$. Thus, in order to find the sender's globally optimal policy, we need to check all possible combinations of $x_i$'s via exhaustive search, and for each possible combination of the $x_i$'s, we need to determine the corresponding optimal values of $s_i^*$ provided in (\ref{eqn:s_i_opt}). During these steps, if there exist solutions with $c_i= c_{i,\min}$ and $s_i=0$, then we can eliminate these solutions directly since they cannot be globally optimal. By this elimination, the computational complexity of the exhaustive search can be reduced and still the global solution will be achieved.             
\vspace{-0.15cm}
\section{Numerical Results}\label{Num_result}
\vspace{-0.15cm}

\begin{figure}[t]
  \centering
  \includegraphics[width=.6\columnwidth]{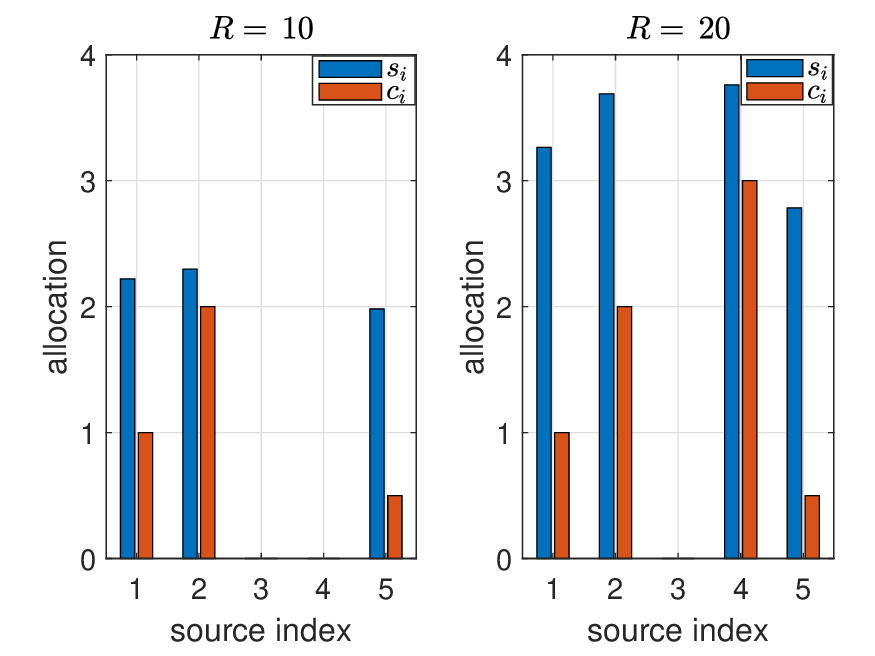}\vspace{-0.35cm}
  \caption{The optimal resource allocation of the sender for $R=\{10,20\}$.}
  \label{fig:AllocationFigure}
  \vspace{-0.25cm}
\end{figure}

In this section, we provide two simulation results to verify the theoretical analysis and highlight key insights. In our first simulation result, we consider 5 different sources with the rates 
\(\lambda=[1.3,\,1.8,\,0.7,\,2.3,\,1.5]\) and 
\(\mu=[2.3,\,3.8,\,3.2,\,5.3,\,2.0]\) and the receiver's utility weight is given by \(q=0.5\). For a given set of active sources with $x_i = 1$, the sender assigns the minimal sampling rates $c_i = c_{i,\min}$ required to satisfy the IC constraint. Under these fixed $c_i$ values, the sender's optimal allocation problem for $s_i$ becomes convex and is expressed in (\ref{eqn:s_i_opt}). By exhaustively searching over all possible sets of active sources, we then determine the optimal $s_i$ allocation and select the set that yields the highest utility for the sender.  We plot the optimal
\((s_i,c_i)\) allocations in Fig.~\ref{fig:AllocationFigure} when the total update rate constraint is $R=\{10,20\}$. For \(R=10\), the optimal active set of sources is
\(\{1,2,5\}\) and the corresponding bars show
positive \(s_i\) only on these indices with \(c_i=c_{i,\min}\). When we increase
the budget to \(R=20\), the sender starts to send updates about source \(4\)'s state and the optimal set becomes
\(\{1,2,4,5\}\).

\begin{figure}[t]
  \centering

  \includegraphics[width=.6\columnwidth]{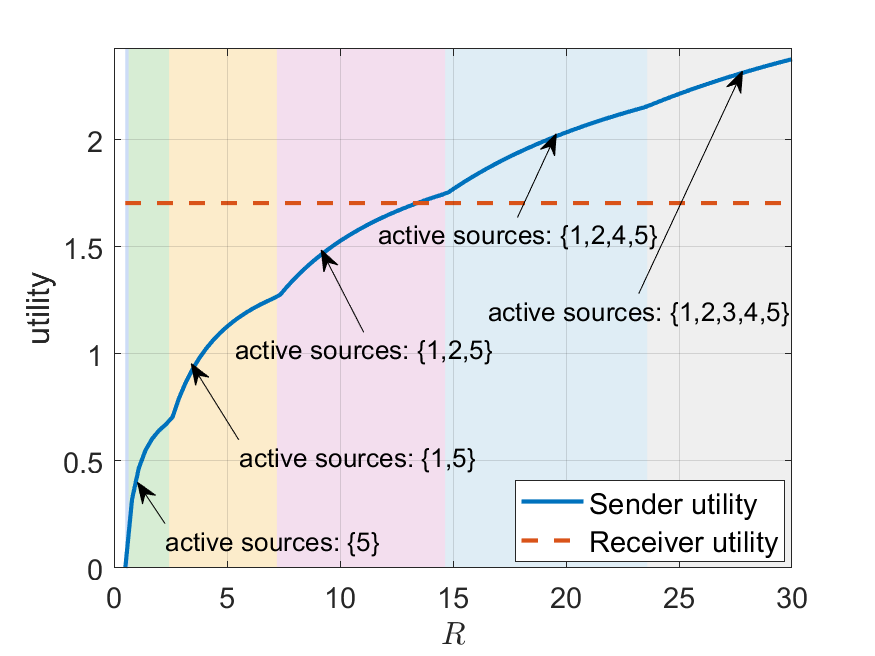}\vspace{-0.45cm}
  \caption{Utility functions  with respect to total budget $R$. } 
  \label{fig:utilityvsR}
  \vspace{-0.05cm}
\end{figure}

Next, we plot the sender’s optimal utility \(J_S^\star(R)\) as the
total budget \(R\) increases, together with the receiver’s benchmark utility which is constant as shown in Fig.~\ref{fig:utilityvsR}. We consider the same set of $\lambda_i$ and $\mu_i$ as before and choose $q=0.5.$ 
The colored bands highlight the different regions which are the intervals of \(R\)  where the
optimal active source set remains the same. For example, when the total sampling rate is very limited, i.e., $0.5\leq R\leq {2.55}$, the sender can only send updates about source 5. As 
$R$ increases, the sender begins to transmit updates about a larger set of sources, specifically in the following order: 
\(\{5\}\), \(\{1,5\}\), \(\{1,2,5\}\), \(\{1,2,4,5\}\), and finally
\(\{1,2,3,4,5\}\). 
Each transition occurs exactly when allocating \(c_{i,\min}\) of the next source and some $s_i$ on that source is more beneficial to the sender compared to allocating budget to the $s_i$ of the active sources. The receiver’s utility curve remains constant by the construction of the problem. We adopt an optimistic assumption that, at the boundary $c_{i,\min}$ where the receiver is indifferent between following the sender’s messages and relying solely on prior information, the receiver chooses to follow the sender’s messages. Consequently, the sender always sets $c_i = c_{i,\min}$ to persuade the receiver, and the receiver attains the same utility under both policies. When we increase the sender's budget $R$, the sender can obtain a higher utility whereas the receiver’s
default utility remains constant across all \(R\).

\begin{figure}[t]
  \centering

  \includegraphics[width=.65\columnwidth]{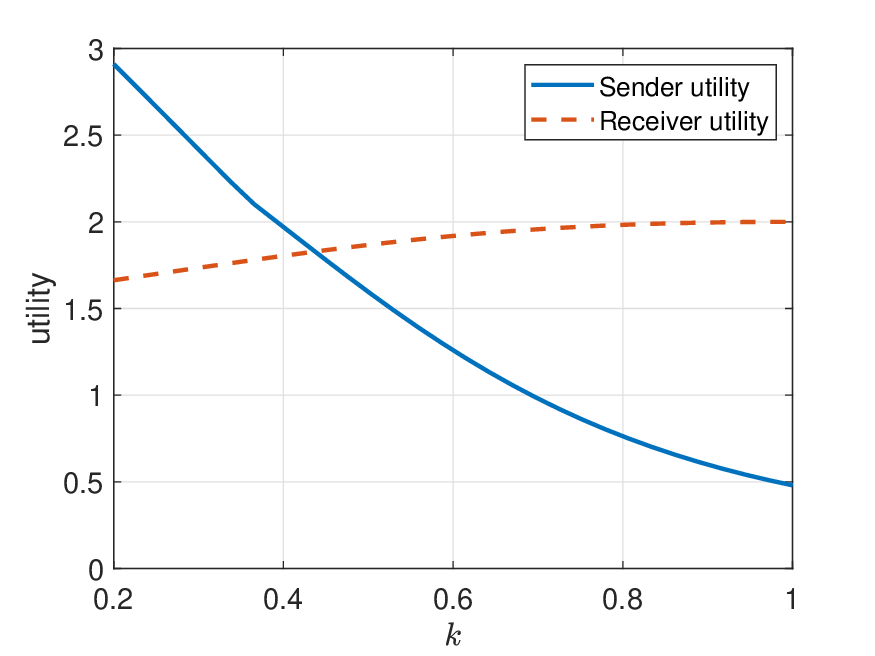}
  \vspace{-0.3cm}
  \caption{The sender's and the receiver's utilities with respect to the heterogeneous distribution of $\mu$.}
  \label{fig:heterok}
   \vspace{-0.3cm}
\end{figure}
In our second simulation result, we choose \(n=5\), \(q=0.5\), \(R=15\), and \(\lambda_i= 1\) for all $i$.
To ensure the positivity of the IC constraint, we impose \(\mu_i\ge 1\).  We generate heterogeneous
\(\mu_i\)-profiles by setting
\begin{align}\nonumber\\[-18pt]
   \mu_i \;=\; 1+\,(C-n)\,\frac{k^i}{\sum_{j=1}^n k^j}, \\[-18pt]\nonumber
\end{align}
so that \(\sum_i \mu_i=C=20\) and each \(\mu_i\ge 1\). We vary the parameter \(k\in[0.2,1]\) to control heterogeneity of the $\mu_i$ distribution. For example, \(k=1\) yields the uniform case where all \(\mu_i\)'s are equal, while
smaller \(k\) values produce a more uneven distribution, with some \(\mu_i\) values close to 1 and others significantly larger.
The blue curve in Fig.~\ref{fig:heterok} shows the optimal sender utility
\(J_S^\star\) as a function of \(k\).  The minimum sender's utility occurs at \(k=1\)
(where \(\mu_i =4\) for all $i$), and the sender's utility increases as \(k\) decreases (more
heterogeneous distribution of $\mu_i$).  Intuitively, with \(q= 0.5\) and \(\lambda_i=1\), the IC
constraint for source \(i\) is \(c_{i,\min}=\mu_i-1\).  More heterogeneous distribution of $\mu_i$ creates some
sources with \(\mu_i\) close to \(1\), hence leads to very small \(c_{i, \min}\) values for some sources. These sources are efficient because they do not require high \(c_{i, \min}\) to activate and also yield high marginal utility for persuasion. As a result, the sender benefits from the heterogeneous distribution of $\mu_i$'s. The red curve shows the receiver’s utility
\(\sum_i \tfrac{q\,\mu_i}{\mu_i+\lambda_i}\). Because \(f(\mu_i)=\mu_i/(\mu_i+1)\) is
\emph{concave} in \(\mu_i\), the receiver's utility (\(q\sum_i  f(\mu_i)\)) is maximized with the uniform $\mu_i$ distribution and
decreases with heterogeneity in $\mu_i$. Thus, the heterogeneity in $\mu_i$ helps the sender (by lowering some \(c_{i,\min}\)) but hurts the receiver, yielding the
opposing trends shown in Fig.~\ref{fig:heterok}.
\vspace{-0.25cm}
\section{Conclusion}
\vspace{-0.25cm}
In this paper, we studied a dynamic variant of a persuasion problem in which a sender influences a receiver by controlling when to reveal information from binary CTMC sources. We derived a closed-form incentive-compatibility constraint for each source, showed that the single-source problem admitted an explicit solution, and formulated the multi-source problem under a total-rate budget. We then designed a globally convergent bisection algorithm computing the optimal state-dependent update rates using the active-set exhaustive search method. Our numerical results demonstrated monotone growth of the sender’s utility with budget, piecewise-constant active sets across budget regions, and predictable shifts under heterogeneous transition rates. As a future research direction, we plan to develop a more efficient algorithm that mitigates the need for the exhaustive solution method presented in Section~\ref{sect:multisourceproblem}. As a further direction, we plan to study multi-sender extensions and hierarchical equilibrium policies under both aligned and misaligned sender objectives.
\vspace{-0.24cm}

\bibliography{refs}

\end{document}